\documentclass[conference]{IEEEtran}
\ifCLASSINFOpdf
\else
\fi

\usepackage{graphicx}
\usepackage[hyphens]{url}
\usepackage{flushend}
\usepackage{amsmath}
\usepackage{subfig}
\usepackage{multirow} 
\usepackage{fixltx2e}

\begin{document}
\title{Early Observations on Performance of Google Compute Engine for Scientific Computing}

\author{\IEEEauthorblockN{Zheng Li}
\IEEEauthorblockA{School of Computer Science\\
ANU and NICTA\\
Canberra, Australia\\
zheng.li@nicta.com.au}
\and
\IEEEauthorblockN{Liam O'Brien}
\IEEEauthorblockA{ICT Innovation and Services\\
Geoscience Australia\\
Canberra, Australia\\
liamob99@hotmail.com}
\and
\IEEEauthorblockN{Rajiv Ranjan}
\IEEEauthorblockA{ICT Center\\
CSIRO and ANU\\
Canberra, Australia\\
raj.ranjan@csiro.au}
\and
\IEEEauthorblockN{Miranda Zhang}
\IEEEauthorblockA{School of Computer Science\\
CSIRO and ANU\\
Canberra, Australia\\
miranda.zhang@csiro.au}}


%Telephone: (800) 555--1212, Fax: (888) 555--1212}


\maketitle

\begin{abstract}
%\boldmath
Although Cloud computing emerged for business applications in industry, public Cloud services have been widely accepted and encouraged for scientific computing in academia. The recently available Google Compute Engine (GCE) is claimed to support high-performance and computationally intensive tasks, while little evaluation studies can be found to reveal GCE's scientific capabilities. Considering that fundamental performance benchmarking is the strategy of early-stage evaluation of new Cloud services, we followed the Cloud Evaluation Experiment Methodology (CEEM) to benchmark GCE and also compare it with Amazon EC2, to help understand the elementary capability of GCE for dealing with scientific problems. The experimental results and analyses show both potential advantages of, and possible threats to applying GCE to scientific computing. For example, compared to Amazon's EC2 service, GCE may better suit applications that require frequent disk operations, while it may not be ready yet for single VM-based parallel computing. Following the same evaluation methodology, different evaluators can replicate and/or supplement this fundamental evaluation of GCE. Based on the fundamental evaluation results, suitable GCE environments can be further established for case studies of solving real science problems.
\end{abstract}


% \fi
\IEEEpeerreviewmaketitle

\section{Introduction}
Cloud computing has emerged originally as a business model \cite{Zhang_Cheng_2010}; therefore, public Cloud services are provided mainly to meet the technological
and economic requirements from business enterprises. However, a previous study shows that Cloud computing is also widely accepted as a potential and encouraging paradigm to solve scientific problems \cite{Li_Zhang_2012}, for the same benefits of on-demand resource provisioning and cost effectiveness. In fact, the existing public Cloud services can be improved for scientific computing through appropriate optimizations \cite{Evangelinos_Hill_2008,Ostermann_Iosup_2009}. On the other hand, once commercial Cloud vendors pay more attention to academic requirements, they can make the current Cloud
more scientific-oriented by slightly changing their infrastructures \cite{He_Zhou_2010}. 
%Interestingly, the industry has started offering highly capable services to help fulfil complex science/engineering tasks \cite{Amazon_2011}.

Driven by the diversity of requirements \cite{Li_Zhang_2012}, Google launched its infrastructure service, namely Google Compute Engine (GCE), and recently made it available to the public \cite{Geelan_2013}. Being ``designed to run high-performance, computationally intensive virtual compute clusters" \cite{Google_instance}, GCE seems to be a promising public Cloud service that can satisfy the requirement of scientific computing. Unfortunately, to the best of our knowledge, no formal GCE evaluation work has been done to date. Although a set of relevant evaluation studies can be found in technical websites, most of them lack comprehensive experiments to reveal GCE's performance, not to mention summarizing GCE's characteristics for scientific computing.
To help understand GCE for dealing with scientific issues, we decided to systematically evaluate GCE following the Cloud Evaluation Experiment Methodology (CEEM) \cite{Li_OBrien_2013}.

%In fact, increasing efforts have been made on evaluating public Cloud services for scientific computing. 
When it comes to evaluating public Cloud services for scientific computing, the efforts can be distinguished between two types according to their evaluation strategies. The first type is to reveal fundamental performance of Cloud services by using traditional and/or scientific benchmarks, which happens usually during the early stage of investigating new Cloud services. For example, data durability, availability and access performance of Amazon S3 were particularly analyzed for supporting science grids \cite{Palankar_Iamnitchi_2008}; and the NAS Parallel Benchmarks (NPB) was used to verify Amazon EC2 for high performance computing \cite{Akioka_Muraoka_2010,Walker_2008}. The second type is to investigate sophisticated Cloud applications as real scientific cases, which is normally based on the aforementioned early-stage evaluation. For example, satellite image processing was studied on Windows Azure \cite{Humphrey_Hill_2011}; and a metabolic flux analysis was implemented by employing multiple types of Amazon services \cite{Dalman_Doernemann_2010}. 

Given the recently available GCE, this paper reports an early-stage performance evaluation. By employing four popular benchmarks, our study exhibits the fundamental performance of four GCE types, and also compares them with nine Amazon EC2 types. Based on the experimental results and analyses, we further show the potential advantages of, and possible threats to applying GCE to scientific computing. For example, GCE would be particularly suitable for applications that require frequent disk operations, while it may not well support single VM-based parallel computing. Following the same evaluation methodology, even different evaluators would be able to replicate and/or supplement this fundamental evaluation of GCE. Moreover, according to the outcome of this study, researchers and engineers can establish suitable GCE environments to carry out sophisticated and scientific case studies. Overall, the contribution of this work is twofold. Firstly, to our best knowledge, this is the first study that systematically reveals GCE's performance for scientific computing. Secondly, this evaluation practice can be viewed as a case study for validating the methodology CEEM.

The remainder of this paper is organized as follows. Section \ref{relatedwork} summarizes the existing practices related to GCE evaluation. Section \ref{methodology} specifies our pre-experimental evaluation activities following the CEEM. The experimental results and analyses around GCE properties including communication, memory, storage, and computation are respectively reported in Section \ref{result}. Conclusions and some future work are discussed in Section~\ref{conclusion}.

\makeatletter 
  \newcommand\figcaption{\def\@captype{figure}\caption} 
\makeatother

\begin{table}[!t]
%% increase table row spacing, adjust to taste
%\renewcommand{\arraystretch}{1.3}
% if using array.sty, it might be a good idea to tweak the value of
% \extrarowheight as needed to properly center the text within the cells
%\caption{The Methodology for Cloud Services Evaluation}
%\label{table_methodology}
\centering
%% Some packages, such as MDW tools, offer better commands for making tables
%% than the plain LaTeX2e tabular which is used here.
\begin{tabular}{|p{8cm}|}
\hline
\begin{enumerate}
\itemindent  -3em
\labelsep 1em
  \item	\textbf{Requirement Recognition:} Recognize the problem, and state the purpose of a proposed evaluation.
  \item	\textbf{Service Feature Identification:} Identify Cloud services and their features to be evaluated.
  \item	\textbf{Metrics and Benchmarks Listing:} List all the metrics and benchmarks that may be used for the proposed evaluation.
  \item	\textbf{Metrics and Benchmarks Selection:} Select suitable metrics and benchmarks for the proposed evaluation.
  \item	\textbf{Experimental Factors Listing:} List all the factors that may be involved in the evaluation experiments.
  \item	\textbf{Experimental Factors Selection:} Select limited factors to study, and also choose levels/ranges of these factors.
  \item	\textbf{Experimental Design:} Design experiments based on the above work. Pilot experiments may also be done in advance to facilitate the experimental design.
  \item	\textbf{Experimental Implementation:} Prepare experimental environment and perform the designed experiments.
  \item	\textbf{Experimental Analysis:} Statistically analyze and interpret the experimental results.
  \item	\textbf{Conclusion and Reporting:} Draw conclusions and report the overall evaluation procedure and results. 
\end{enumerate}\\
\hline
\end{tabular}
\figcaption{\label{table_methodology}CEEM for Cloud services evaluation (cf.~\cite{Li_OBrien_2013}).}
\end{table}

\section{Related Work}
\label{relatedwork}
As revealed in \cite{Li_Zhang_2012}, the requirements in the Cloud market are diverse; Infrastructure as a Service (IaaS, e.g., Amazon EC2) and Platform as a Service (PaaS, e.g., Google AppEngine) would then serve different types of requirements, and they cannot be replaced with each other. This could be the motive for Google to finally launch its infrastructure service GCE \cite{Geelan_2013}. Although GCE has been widely viewed as a strong competitor against the other IaaS services, especially against Amazon EC2, not much evaluation work can be found for understanding GCE's performance. The very early GCE evaluation studies tended to be qualitative discussions around service level agreement (SLA) \cite{Emison_2012,Wayner_2012}. More quantitative evaluations appeared in the latter GCE studies. For example, Zencoder compared GCE's n1-standard-8-d with two Amazon cluster instance types for video transcoding \cite{Zencoder_2012}. As argued in \cite{Emison_2012b}, however, this work did not reveal suitable EC2 types for ``apple-to-apple" comparison to GCE. On the contrary, it is able to identify comparable EC2 types for GCE n1-standard-4 in a serial of evaluation experiments \cite{Murphy_2013,Murphy_2013b}. Nevertheless, as suggested by the author himself, those ad hoc and short-time experiments could result in biases and misleading conclusions. A various-type virtual machine (VM) evaluation of GCE and EC2 was given in \cite{Stadil_2013}. However, the non-standard benchmarks and unclear evaluation procedure make this study hard for experimental replication and comparison. Given the lack of a systematic GCE evaluation, we use this paper to report our empirical study to help understand the fundamental performance of GCE, particularly for scientific computing.

\section{Evaluation Methodology}
\label{methodology}
To achieve convincing evaluation results, we followed the ten-step CEEM \cite{Li_OBrien_2013} to evaluate the selected GCE and EC2 instances (cf.~\textit{VM Type} in \ref{factor}), as illustrated in Fig.~\ref{table_methodology}. By providing systematic guidelines together with evaluation experiences, CEEM is supposed to help reduce human bias and facilitate implementations of Cloud services evaluation. Here we briefly introduce a set of pre-experimental evaluation activities instructed by CEEM.

\subsection{Requirement Recognition and Service Feature Identification}
As mentioned previously, this work is to investigate the fundamental performance of GCE for scientific computing. Since the GCE service is offered as virtual machine (VM) instances, four main characteristics (network I/O, memory, storage, and processing capabilities) can be used to distinguish between different instance types \cite{Murphy_2013}. By using the taxonomy of Cloud services evaluation \cite{Li_OBrien_2012a}, we formally identify the requirement and service features to be evaluated as: \textit{Given particular benchmarks, how capable are GCE instances in terms of the following properties?} Note that a service feature is defined as a combination of a service property and its capability \cite{Li_OBrien_2012a}.
\begin{itemize}
\renewcommand{\labelitemi}{$\bullet$}
\itemsep 2pt
    \item	Communication
    \item	Memory
    \item	Storage
    \item	Computation
\end{itemize}

\subsection{Metrics/Benchmarks Listing and Selection}
Available metrics and benchmarks for Cloud services evaluation can be conveniently explored in the existing metric catalogue \cite{Li_OBrien_2012b}. According to the past evaluation experiences, we decided to select relatively lightweight and popular benchmarks recommended in \cite{Li_Zhang_2012}, as listed in Table \ref{table_benchmark}. In detail, Iperf is able to deliver more precise results by consuming less system resources \cite{Schad_Dittrich_2010}, which is usually employed together with Ping for communication evaluation; the sequential disk operations simulated by Bonnie++ are more typical to scientific computing \cite{Ostermann_Iosup_2009}; STREAM is the de facto memory evaluation benchmark included in the HPC Challenge Benchmark (HPCC) suite \cite{Iosup_Ostermann_2011}; while the NPB suite (including eight benchmarks: BT, CG, EP, FT, IS, LU, MG, and SP) has been widely adopted in the early-stage evaluation of Cloud services for high performance computing \cite{Akioka_Muraoka_2010,He_Zhou_2010,Walker_2008}. In particular, considering LU is the most process-flexible pseudo application benchmark in the suite \cite{NASA_2013}, we chose the benchmarking result of LU to calculate the computation performance/price ratio for different VM types. 

\begin{table}[!t]
%% increase table row spacing, adjust to taste
\renewcommand{\arraystretch}{1.3}
% if using array.sty, it might be a good idea to tweak the value of
% \extrarowheight as needed to properly center the text within the cells
\caption{Metrics and Benchmarks for Evaluating GCE}
\label{table_benchmark}
\centering
%% Some packages, such as MDW tools, offer better commands for making tables
%% than the plain LaTeX2e tabular which is used here.
\begin{tabular}{|l|c|c|c|}
\hline
\textbf{Service Property} & \textbf{Capability Metric} & \textbf{Benchmark} & \textbf{Version}\\
\hline
Communication & Data Throughput & Iperf & 2.0.5 \\
\hline
Communication & Latency & Ping & N/A\\
\hline
Memory & Data Throughput & STREAM & 5.10\\
\hline
Storage & Transaction Speed & Bonnie++ & 1.96\\
\hline
Storage & Data Throughput & Bonnie++ & 1.96\\
\hline
Computation & Transaction Speed & NPB-MPI & 3.3\\
\hline
Computation & Performance/Price Ratio & LU in NPB-MPI & 3.3\\
\hline
\end{tabular}
\end{table}

\subsection{Experimental Factor Listing and Selection}
\label{factor}
Similarly, candidate factors that may influence the measured performance of GCE can be identified within an experimental factor framework \cite{Li_OBrien_2012c}. In general, we considered at least two input factors and one output factor in the evaluation of every service property, as specified below.

\begin{table}[!t]
\renewcommand{\arraystretch}{1.3}
\caption{GCE VM Types Involved in this Evaluation}
\label{table_vmtype}
\centering
\begin{tabular}{|l|c|c|c|c|}
\hline
\textbf{VM Name} & \textbf{Virtual CPUs}& \textbf{GCEUs}  & \textbf{Memory} & \textbf{Scratch Disk}\\
\hline
n1-standard-1-d & 1 & 2.75 & 3.75GB & 420GB \\
\hline
n1-standard-2-d & 2 & 5.5 & 7.5GB & 870GB\\
\hline
n1-highmem-2-d & 2 & 5.5 & 13GB & 870GB\\
\hline
n1-highcpu-2-d & 2 & 5.5 & 1.8GB & 870GB\\
\hline
n1-standard-1 & 1 & 2.75 & 3.75GB & 0\\
\hline
\textbf{Image} & \multicolumn{4}{c|}{SCSI-enabled GCEL 12.04 LTS (gcel-12-04-v20130104)}\\ 
\hline
\end{tabular}
\end{table}

\begin{itemize}
\renewcommand{\labelitemi}{$\bullet$}
\itemsep 2pt
    \item	\textit{VM Type:} At the time of writing, Google supplies 10 types of VM instances (with local disks) categorized into three groups \cite{Google_instance}. Given the limitation of quota (8 CPUs) \cite{Google_quota}, it is impossible to evaluate all the VM types during the same experimental time window. Therefore, we chose four types (the first four in Table \ref{table_vmtype}) as the variable values of \textit{VM Type} to cover the three instance groups for memory, storage and computation evaluation. With regard to the communication evaluation, we only chose the type n1-standard-1 for the consistent environment and the lowest price. Note that, for the purpose of comparison, we also evaluated nine 64bit-Linux-image EC2 types located in Amazon's North Virginia data center (us-east-1b zone). Due to the limit of space, we directly report the EC2-related experimental results without listing the individual EC2 types here.
    \item	\textit{Duration:} For each evaluation experiment, we decided to take a whole-day observation including one-hour warming up. In other words, we assign the value of \textit{Duration} as 24 hours.
    \item	\textit{Capability Metric:} The corresponding capability metric of each service property can be viewed as an output factor or a response \cite{Li_OBrien_2012c}, as shown in Table \ref{table_benchmark}. By employing two different benchmarks, we essentially considered two output factors of the communication property. 
\end{itemize}

In particular, we supplemented three more input factors to communication evaluation.
\begin{itemize}
\renewcommand{\labelitemi}{$\bullet$}
\itemsep 2pt
    \item	\textit{(GCE) Geographical Location:} Currently GCE service has five available zones within three geographical locations (i.e.~us-central1, us-central2 and europe-west1). Due to the quota limitation again, we only selected three *-a zones as the VM-side locations.   
    \item	\textit{(Client) Geographical Location:} We varied the client-side locations by using a local machine in our NICTA Canberra Lab and an EC2 micro instance in Amazon's us-east-1b zone.
    \item	\textit{Communication Scope:} Given the two communication patterns, intra-Cloud and wide-area \cite{Li_OBrien_2012c}, we decided to respectively observe the data transferring performance between GCE instances and between a GCE instance and a client. Note that, in all the communication experiments, we only visited the external IP addresses of GCE instances to force their network address translation (NAT). As such, the result of the intra-Cloud evaluation would be more comparable to the wide-area scenario. %the communication inside a GCE location can still be viewed as a very close client-Cloud scenario.
\end{itemize}

\begin{table}[!t]
%% increase table row spacing, adjust to taste
\renewcommand{\arraystretch}{1.3}
% if using array.sty, it might be a good idea to tweak the value of
% \extrarowheight as needed to properly center the text within the cells
\caption{Experimental Design for Communication Evaluation}
\label{table_communication}
\centering
%% Some packages, such as MDW tools, offer better commands for making tables
%% than the plain LaTeX2e tabular which is used here.
\begin{tabular}{|l|c|c|c|}
\hline
\textbf{Locations} & GCE-central1-a & GCE-central2-a & GCE-west1-a\\
\hline
Local & X & X & X \\
\hline
EC2-us-east-1b & X & X & X\\
\hline
GCE-central1-a & X & X & X\\
\hline
GCE-central2-a &  & X & X\\
\hline
GCE-west1-a &  &  & X\\
\hline
\end{tabular}
\end{table}

When it comes to computation evaluation, we were further concerned with \textit{Workload Size} and \textit{Process Number} as two experimental factors.
\begin{itemize}
\renewcommand{\labelitemi}{$\bullet$}
\itemsep 2pt
    \item	\textit{Workload Size:} It can be found that NPB Class A and B are two popular workload sizes in the relevant Cloud benchmarking results \cite{CloudHarmony_2013,Walker_2008}. To make our study easily comparable with the others, we also chose the widely-used workload Class A and B when running the NPB-MPI benchmarks.
    \item	\textit{Process Number:} To decide the number of processes for experimental design, we were concerned with two sides. From the side of resource, as specified in \cite{Google_instance}, the selected GCE instance types have either one or two virtual CPU cores. From the side of workload, although six benchmarks in the NPB-MPI suite run on a power-of-2 number of processes (1, 2, 4, ...), BT and SP run on a square number of processes (1, 4, 9, ...) \cite{NASA_2013}. To satisfy the process variance of all the benchmarks, we naturally distinguished between requesting one and four processes for running BT and SP, and requesting one, two and four processes for running the other NPB-MPI benchmarks. 
\end{itemize}

\begin{figure*}[!t]
\centering
\includegraphics{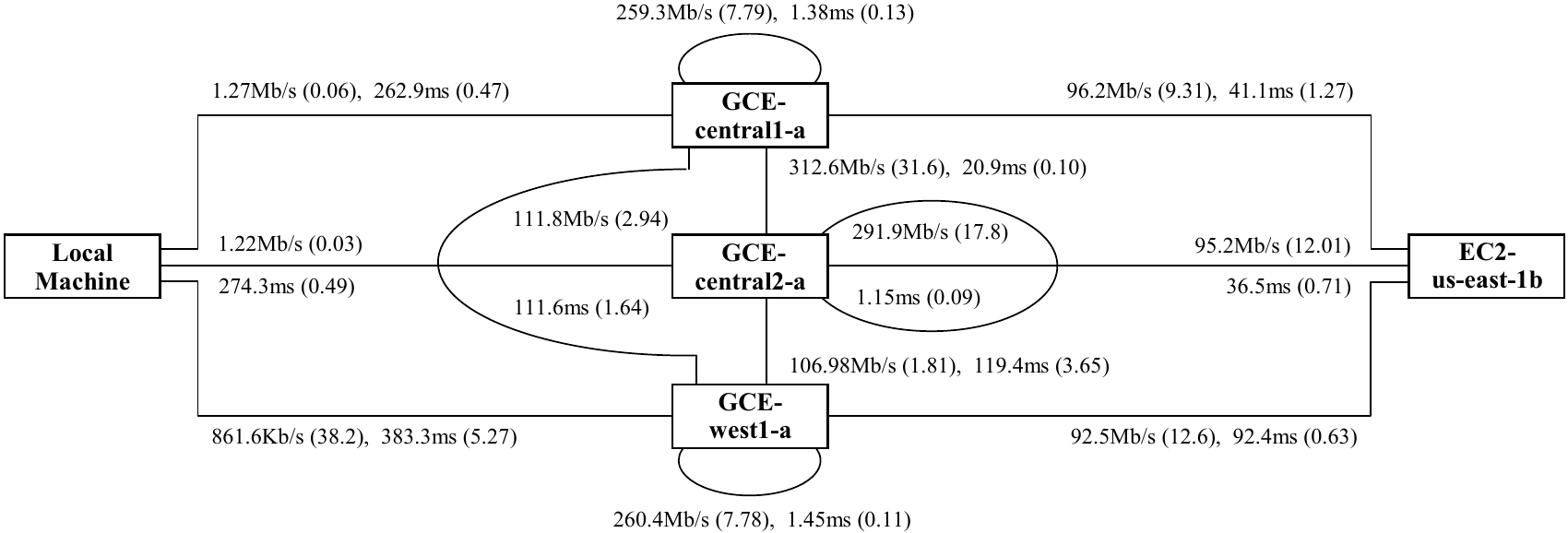}
\caption{Communication benchmarking results by using Iperf and Ping. Each link represents the communication inside one or between two locations, and shows its average data throughput (standard deviation) and average round-trip latency (standard deviation).}
\label{fig>PicCommunication}
\end{figure*}

\subsection{Experimental Design}
By using Design of Experiment (DOE) techniques \cite{Montgomery_2009}, the evaluation experiments can be prepared based on the pre-identified factors. Recall that \textit{Duration} has been constrained to be a single value (i.e.~24 hours), a simple and general design is to run each benchmark on different \textit{VM Type} instances independently for a whole day plus one hour. When it comes to a combination of input factors with various values, more sophisticated experimental design could be required, as specified below.

In the case of communication evaluation that combines \textit{Geographical Location} with \textit{Communication Scope}, the intra-Cloud pattern has six paths ($=C(3,1)+C(2,1)+C(1,1)$, e.g., ``GCE-central1-a to GCE-west1-a" and ``GCE-west1-a to GCE-central1-a" are treated as the same path) among three server-side locations, and the wide-area pattern also has six paths (=$C(3,1){\times}C(2,1)$) between three server-side and two client-side locations. The design result is directly listed in Table \ref{table_communication}, where an \textit{X} indicates a communication path that needs to be evaluated. 

As for the computation evaluation that combines \textit{Workload Size} and \textit{Process Number}, we employed the Full-factorial Design technique \cite{Montgomery_2009} to statistically investigate those factors' influences on the performance of GCE instances. This design technique adjusts one factor at a time. Also recall that there are eight individual benchmarks with two patterns of process variance in the NPB-MPI suite; a single round of experiment on a VM then consists of 44 (=$2{\times}2{\times}2+2{\times}3{\times}6$) different benchmarking trials.

\section{Evaluation Results and Analyses}
\label{result}
\subsection{Communication Evaluation Result and Analysis}

To save space, we show the 12-path communication evaluation results together in Fig.~\ref{fig>PicCommunication}. In general, the data throughputs of communication within US GCE data centers are approximately 2.5 times higher than that between US and Europe, while the lowest one is around 107Mb/s between GCE-central2 and GCE-west1. Recall that NAT is enforced by visiting only external IP addresses in this communication evaluation. If using internal IP addresses, according to our ad hoc experiments, we may achieve nearly double of the communication performance inside or between GCE data centers. Such a level of communication data throughput implies that GCE employs the 1000Mb/s Ethernet. 

In particular, the communication performance from NICTA Canberra Lab (also 1000Mb/s Ethernet) to GCE data centers are much worse compared to the measurement from Amazon EC2. This could be caused of much longer routing traces over the connection from our side. Similar to the communication between GCE's US and European data centers, there is also a significant performance decrease when visiting GCE-west1 from the local machine. However, we did not observe this clear trend when measuring from the EC2 instance. Surprisingly, the communication between EC2-us-east and GCE-west1 even has shorter round-trip latency than that between GCE-centrals and GCE-west1. Another surprising observation is that the communication data throughput between GCE-central1 and GCE-central2 is averagely higher than inside the GCE data centers, although it also shows the most variation.

\subsection{Memory Evaluation Result and Analysis}

\begin{figure}[!t]
\centering
\includegraphics[width=8.5cm]{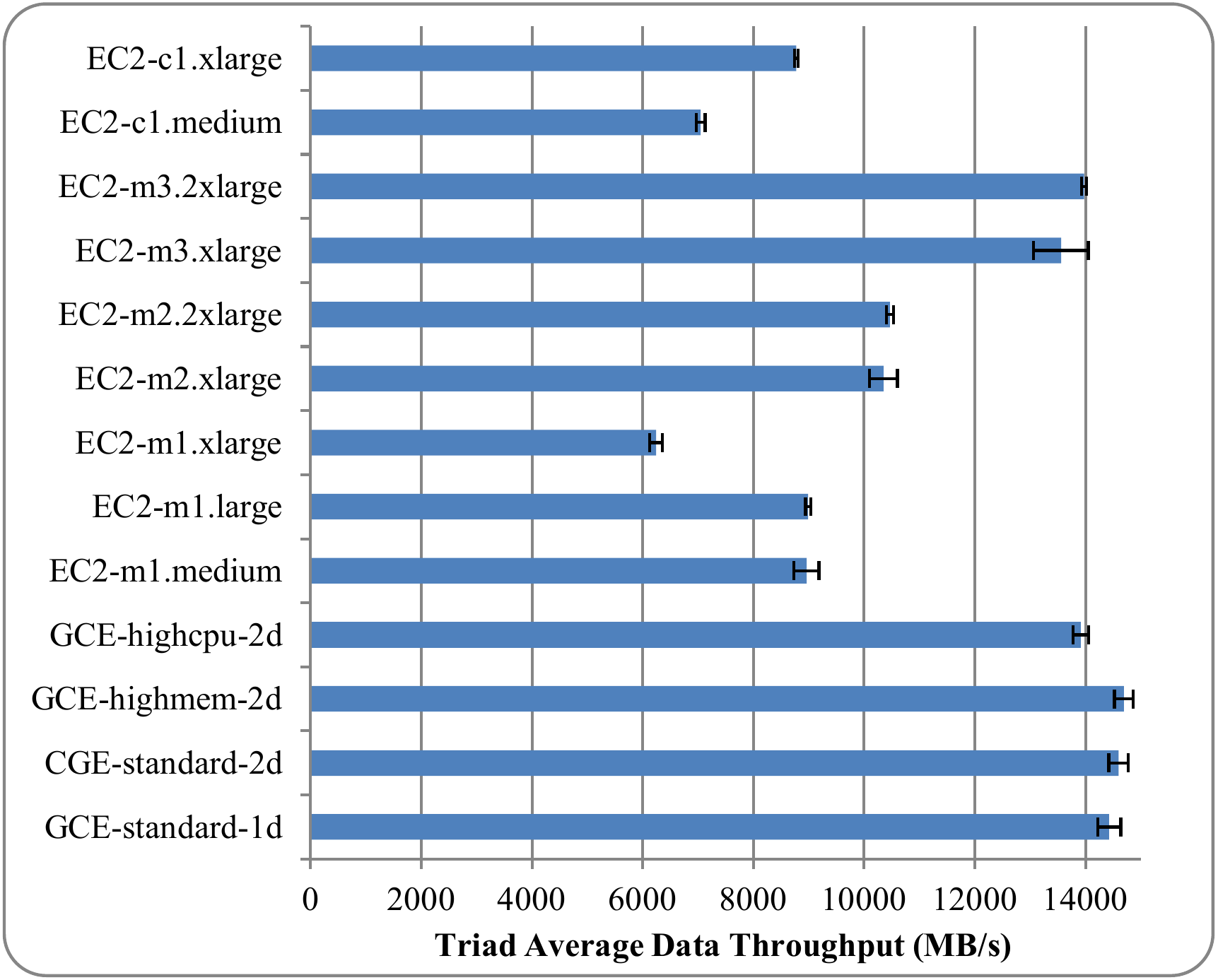}
\caption{Memory benchmarking results (Triad only) by using STREAM. Error bars indicate the standard deviations of the corresponding memory data throughput.}
\label{fig>PicSTREAM}
\end{figure}

As mentioned previously, we also evaluated nine EC2 instances from four type families for comparison to the GCE instances. Corresponding to the single-CPU GCE type (i.e. n1-standard-1-d), we only show the single-thread memory performance of different VM types. As illustrated in Fig.~\ref{fig>PicSTREAM}, the GCE instances displayed stronger memory performance against most of the evaluated EC2 instances. Only Amazon's second generation (M3) standard family can relatively come up to the four GCE types, as specified in Table \ref{table_stream}. This observation is also consistent with the study \cite{Murphy_2013b}, although the author evaluated a different GCE type (i.e. n1-standard-4) from this work. In addition, interestingly, EC2 m3.2xlarge seems to be one of the most memory-stable VM types, while EC2 m3.xlarge exposed the most variability in terms of memory data throughput.

Furthermore, it seems that Google adopts different memory virtualization strategy from Amazon. Fig.~\ref{fig>PicSTREAM} shows that Amazon varies memory capabilities for different families of VM types. Even in the same type family, Amazon may have deliberately decreased the memory data throughput for higher-SLA VM instances (e.g., EC2 m1.xlarge in the first generation (M1) standard family). On the contrary, this work along with the related study \cite{Murphy_2013b} show that Google may offer relatively consistent memory performance for all the GCE instance types. As such, only take memory into account, Google supplies an easier and clearer condition for VM type selection. Moreover, since even the cheapest GCE instance (n1-standard-1-d in this case) shows high memory performance, GCE service may be considered as an economic candidate for memory-intensive computing.

\begin{table}[!t]
\renewcommand{\arraystretch}{1.3}
\caption{Several Comparable Memory Performance}
\label{table_stream}
\centering
\begin{tabular}{|l|c|c|c|c|}
\hline
\multirow{2}{*}{\textbf{VM Type}} & \textbf{Copy(MB/s)}& \textbf{Scale(MB/s)}  & \textbf{Add(MB/s)} & \textbf{Triad(MB/s)}\\
& \textbf{(Std. Dev.)}& \textbf{(Std. Dev.)}  & \textbf{(Std. Dev.)} & \textbf{(Std. Dev.)}\\

\hline
\multirow{2}{*}{n1-standard-1-d} & 13260.06 & 13098.5 & 14863.63 & 14430.51\\
 & (190.3) & (185.21) & (220.18)& (208.18)\\
\hline
\multirow{2}{*}{n1-standard-2-d} & 13420.94 & 13274.56 & 15038.48 & 14597.56\\
&(156.7)&(146.25)&(170.23)&(176)\\
\hline
\multirow{2}{*}{n1-highmem-2-d} & 13472.22&13332.92&15113.48&14694.01\\
 &(148.01)&(145.15)&(171.8)&(168.47)\\
\hline
\multirow{2}{*}{n1-highcpu-2-d} & 12795.99&12668.99&14328.06&13919.97\\
&(128.34)&(135.68)&(136.28)&(140.82)\\
\hline
\multirow{2}{*}{m3.xlarge} & 12575.48 & 12264.63 & 13891.32 & 13559.57\\
& (540.21) & (518.3) & (524.82) & (494.86)\\
\hline
\multirow{2}{*}{m3.2xlarge} & 13044.89 & 12722.88 & 14341.53 & 13974.84\\
 & (50.05) & (39.76) & (59.89) & (43.46)\\
\hline
\end{tabular}
\end{table}

\begin{figure}
  \centering
  \subfloat[Storage transaction speed.]{
    \label{fig:subfigBonnieChar} %% label for first subfigure
    \includegraphics[width=8.5cm]{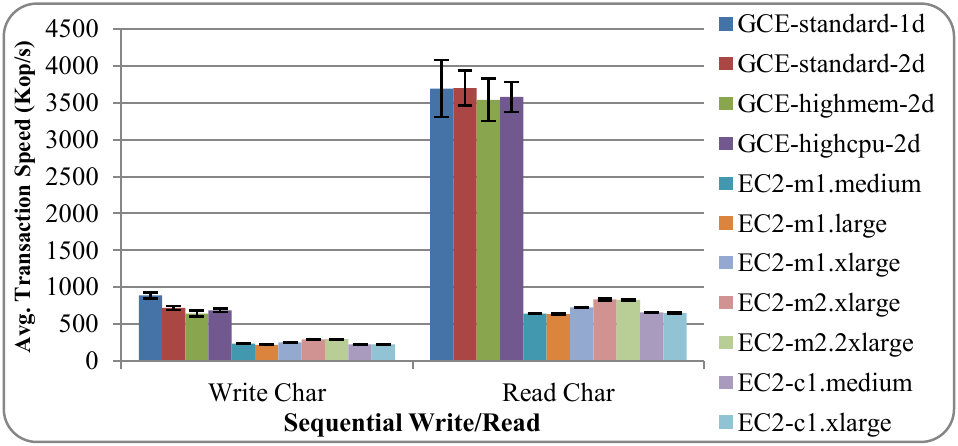}}

  \subfloat[Storage data throughput.]{
    \label{fig:subfigBonnieBlock} %% label for second subfigure
    \includegraphics[width=8.5cm]{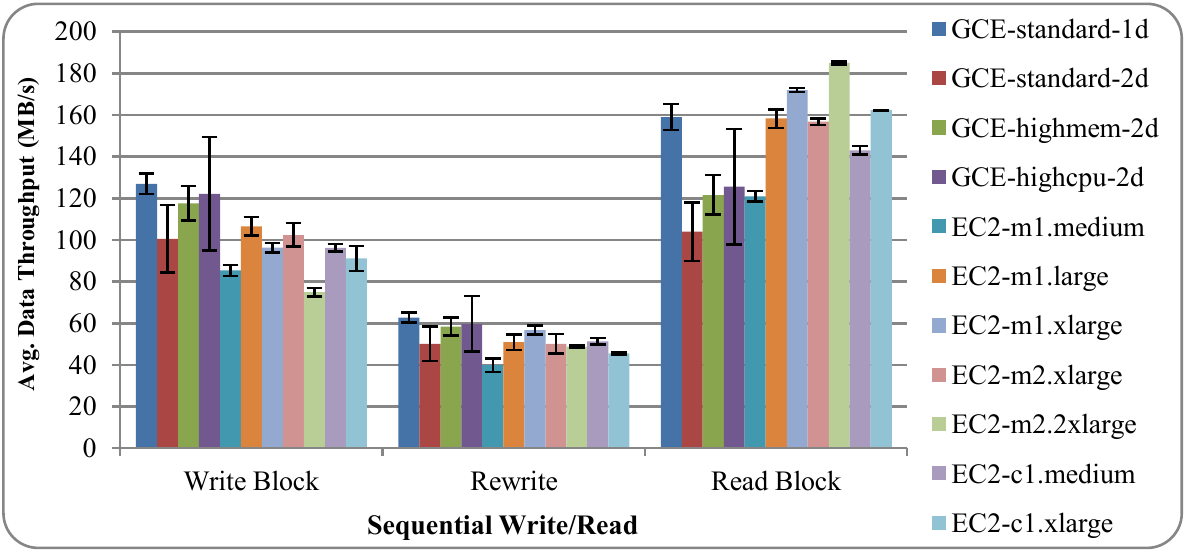}}
  \caption{Storage benchmarking results using Bonnie++. Error bars indicate the standard deviations of the corresponding storage performance.}
  \label{fig:subfigBonnie} %% label for entire figure
\end{figure}

\begin{figure}[!ht]
  \centering
  \subfloat[GCE n1-standard-1-d.]{
    \label{fig:subfig:a} %% label for first subfigure
    \includegraphics[width=4.2cm]{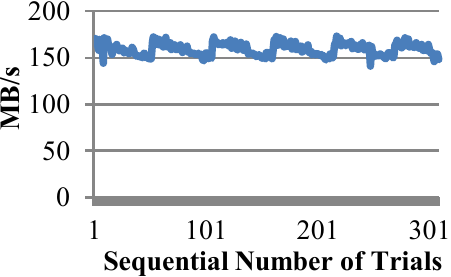}}
  \subfloat[GCE n1-standard-2-d.]{
    \label{fig:subfig:b} %% label for second subfigure
    \includegraphics[width=4.2cm]{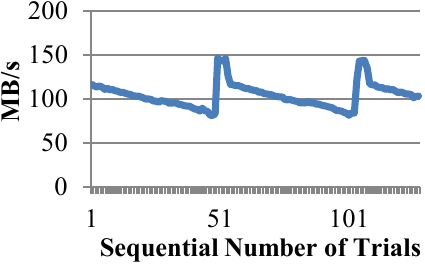}}

  \subfloat[GCE n1-highmem-2-d.]{
    \label{fig:subfig:c} %% label for first subfigure
    \includegraphics[width=4.2cm]{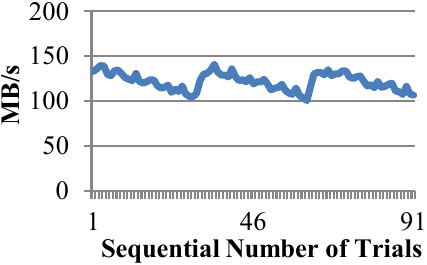}}
  \subfloat[GCE n1-highcpu-2-d.]{
    \label{fig:subfig:d} %% label for second subfigure
    \includegraphics[width=4.2cm]{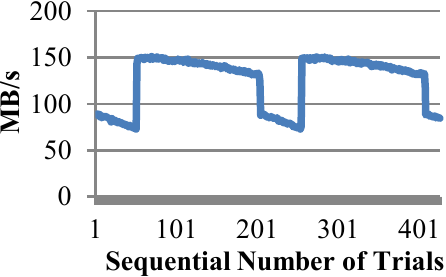}}

  \subfloat[EC2 m1.large.]{
    \label{fig:subfig:e} %% label for first subfigure
    \includegraphics[width=4.2cm]{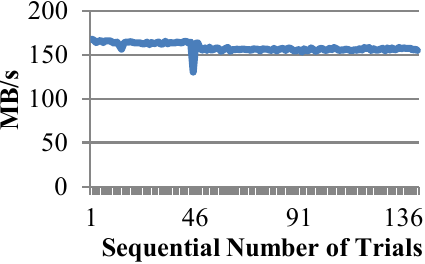}}
  \subfloat[EC2 m2.2xlarge.]{
    \label{fig:subfig:f} %% label for second subfigure
    \includegraphics[width=4.2cm]{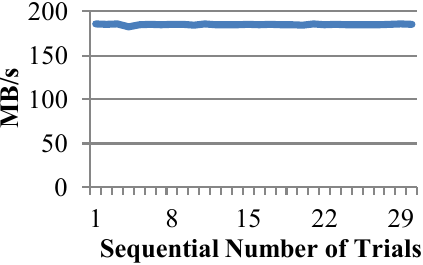}}
  \caption{Block reading using Bonnie++ during a whole day.}
  \label{fig:subfig} %% label for entire figure
\end{figure}

\begin{figure*}
  \centering
  \subfloat[GCE n1-standard-1-d.]{
    \label{fig:subfigComputation:a} %% label for first subfigure
    \includegraphics[width=8.4cm]{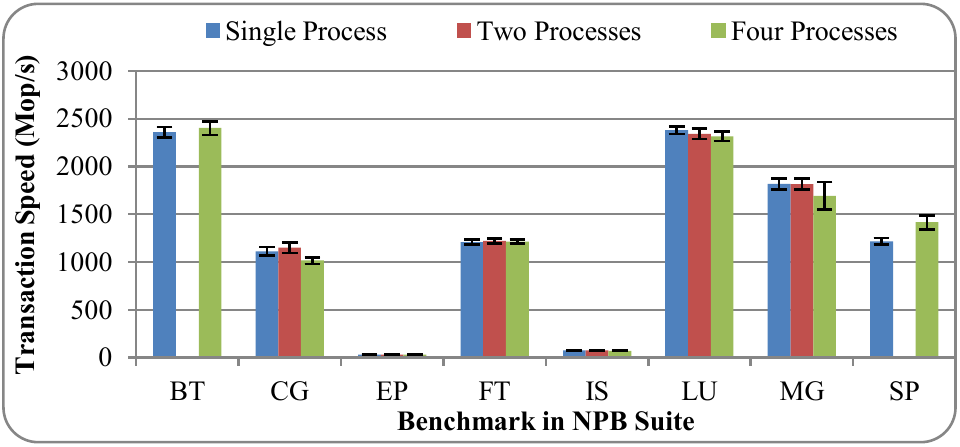}}
  \subfloat[GCE n1-standard-2-d.]{
    \label{fig:subfigComputation:b} %% label for second subfigure
    \includegraphics[width=8.4cm]{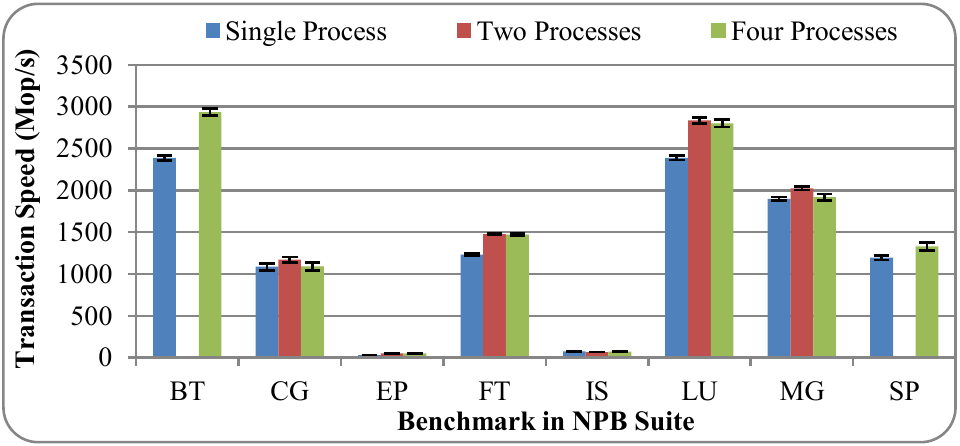}}
  \hspace{0in}
  \subfloat[GCE n1-highmem-2-d.]{
    \label{fig:subfigComputation:c} %% label for first subfigure
    \includegraphics[width=8.4cm]{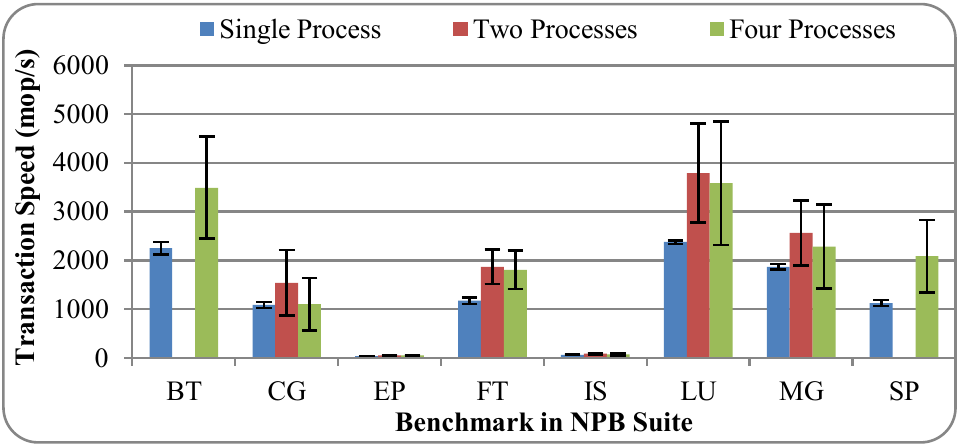}}
  \subfloat[GCE n1-highcpu-2-d.]{
    \label{fig:subfigComputation:d} %% label for second subfigure
    \includegraphics[width=8.4cm]{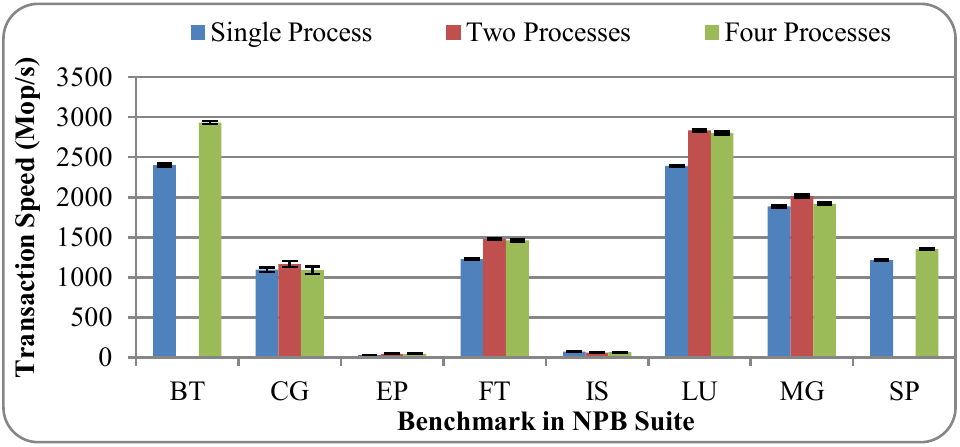}}
  \hspace{0in}
  \subfloat[EC2 m1.large.]{
    \label{fig:subfigComputation:e} %% label for first subfigure
    \includegraphics[width=8.4cm]{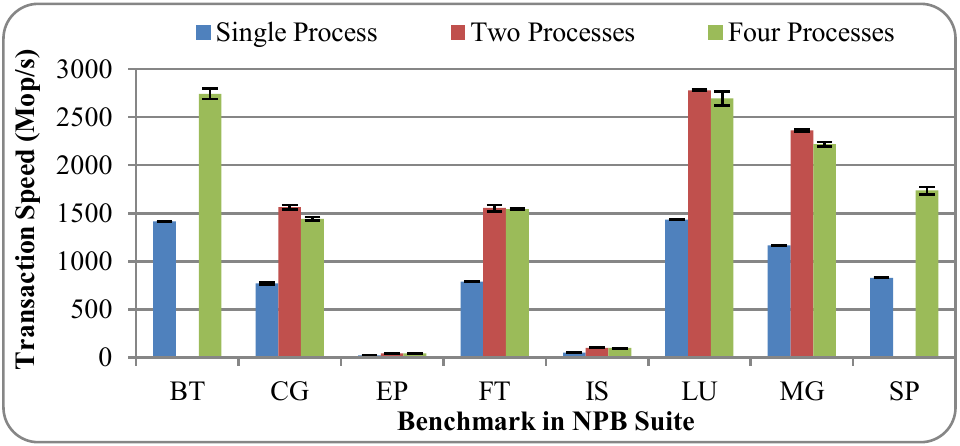}}
  \subfloat[EC2 m3.xlarge.]{
    \label{fig:subfigComputation:f} %% label for second subfigure
    \includegraphics[width=8.4cm]{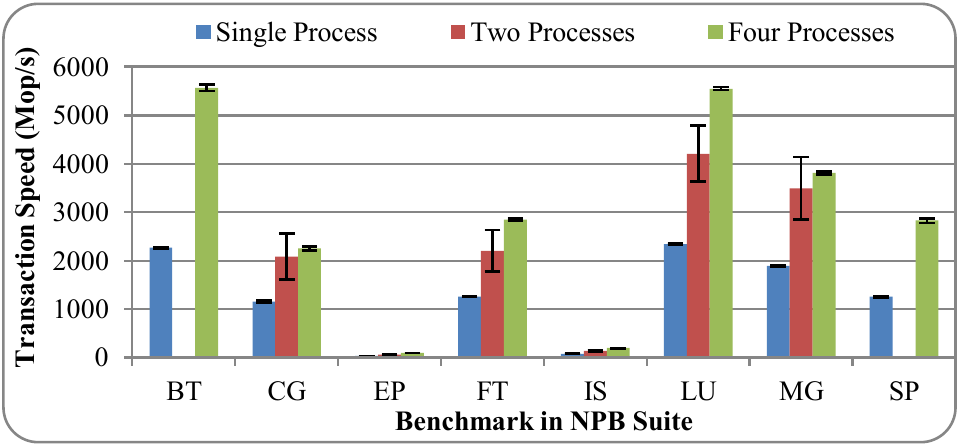}}
  \caption{Computation benchmarking results using NPB-MPI with workload Class A. Error bars indicate the standard deviations of the corresponding computation transaction speed.}
  \label{fig:subfigComputation} %% label for entire figure
\end{figure*}

\subsection{Storage Evaluation Result and Analysis}

%\begin{figure}[!t]
%\centering
%\includegraphics[width=8.5cm]{PicBonnieChar.pdf}
%\caption{Storage benchmarking results (transaction speed) using Bonnie++. Error bars indicate the standard deviations of the corresponding storage transaction speed.}
%\label{fig>PicBonnieChar}
%\end{figure}

%\begin{figure}[!t]
%\centering
%\includegraphics[width=8.5cm]{PicBonnieBlock.pdf}
%\caption{Storage benchmarking results (data throughput) using Bonnie++. Error bars indicate the standard deviations of the corresponding storage data throughput.}
%\label{fig>PicBonnieBlock}
%\end{figure}

% An example of a floating figure using the graphicx package.

It has been identified that sequential disk operations are more typical to scientific computing \cite{Ostermann_Iosup_2009}. To save space, here we only show the sequential storage performance of the evaluated VM types. In particular, we did not use Bonnie++ to benchmark Amazon's second generation (M3) standard instances. The M3 instances employ persistent disks of Elastic Block Store (EBS) only. Since we did not consider GCE's persistent storage in this study, EBS evaluation is then out of the scope of this paper.

Firstly, we assessed the storage transaction speed of each instance with sequential write/read per character. When it comes to character write/read, in fact, Bonnie++ only measures the amount of data performed per second. Considering that each byte of data incurs a transaction in this case, we can directly translate Bonnie++'s result into storage transaction speed, as shown in Fig.~\ref{fig:subfigBonnieChar}. Among the evaluated VM types, it is clear that GCE has overwhelming advantage over EC2 on writing/reading small size of data, although there is considerable variability in reading on GCE instances. Given the approximately 3- to 4.5-times faster storage transaction speed, GCE could be particularly suitable for applications that need frequent disk access.

Secondly, we assessed the storage data throughput of each instance with sequential write/read per block. Interestingly, the benchmarking result as illustrated in Fig.~\ref{fig:subfigBonnieBlock} shows opposite performance trends in block writing and reading: GCE instances seem better at writing in general, while EC2 instances generally win at reading. Moreover, unlike EC2, individual GCE instance's (except for n1-standard-1-d's) reading and writing data throughputs are nearly the same. A possible explanation for this is that Amazon has implemented storage cache for EC2, while GCE seems not to be the case.

Similar to accessing small size of data, block writing/reading on GCE instances also displays considerable variance. To better observe the storage variability of different VM instances, we plotted the benchmarking results against trial sequence of the whole day. Note that Bonnie++ generates test files that are at least twice the memory size of the evaluated machine to invalidate the memory's interference, while bigger test files inevitably take longer time in a trial. Therefore, the number of trials during a whole day would be different for different memory-size VM types. Surprisingly, all the four GCE types have regular patterns of performance jitter in block writing, rewriting and reading. This paper only reports their block reading patterns that are the clearest, as illustrated in Fig.~\ref{fig:subfig}. As we know, the hard disk surface is divided into a set of concentrically circular tracks. Given the same rotational speed of a hard disk drive (HDD), the outer tracks would have higher data throughput than the inner ones. As such, the regular patterns may indicate that the HDD heads sequentially shuttle between outer and inner tracks when consecutively writing/reading block data on GCE instances.

\begin{figure*}[!t]
\centering
\includegraphics{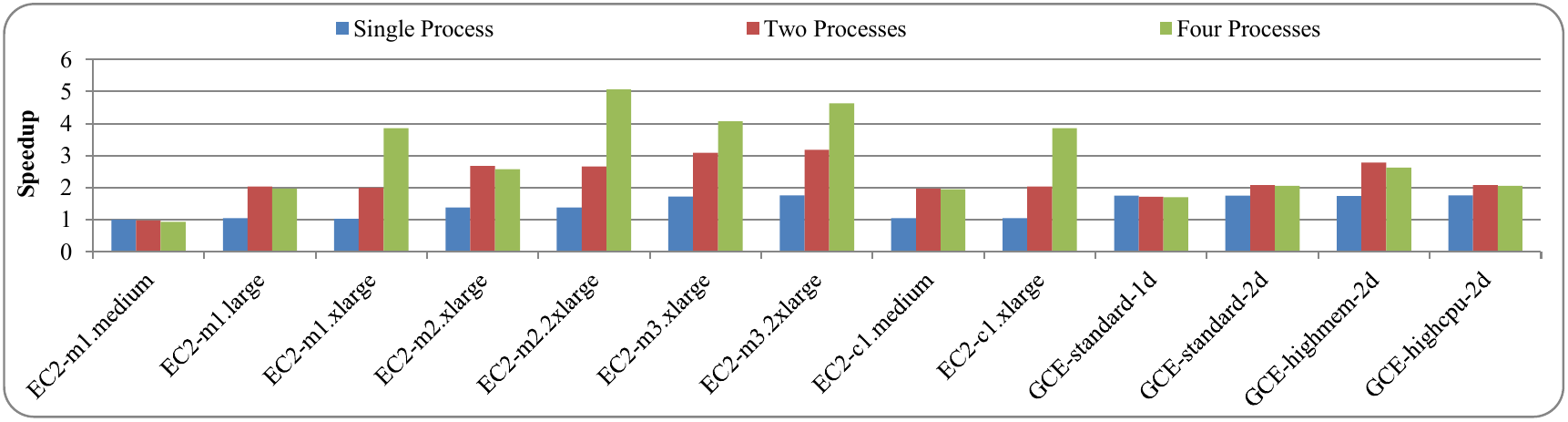}
\caption{Computation speedup over the single-process performance of EC2 m1.medium using LU in the NPB suite. }
\label{fig>PicSpeedup}
\end{figure*}

For comparison, we additionally show the plots of two EC2 instance types. In detail, the m1.large instance had the most variable block data throughput among those EC2 VMs (cf.~Fig.~\ref{fig:subfig:e}), while some tiny and regular jitters in block reading also appeared on the m2.2xlarge instance (cf.~Fig.~\ref{fig:subfig:f}). However, contrasted with GCE, there is no clearly fluctuating pattern on EC2 instances. This phenomenon may confirm the aforementioned hypothesis that Amazon have employed storage cache to iron the block writing/reading fluctuations. Overall, this study verifies that EC2 supplies more stable storage service than GCE.

\subsection{Computation Evaluation Result and Analysis}

Given the literature investigation \cite{Li_Zhang_2012}, NPB could be one of the most well-known and widely-accepted benchmark suites for early-stage scientific computing evaluation. In fact, the NPB suite can cover various Cloud service properties in addition to computation, for example, BT is a disk I/O-intensive benchmark while MG is communication-intensive. In this study, however, we use NPB to emphasize computation only. In other words, the eight NPB benchmarks are employed together to reflect the overall computation performance of a particular VM type.

After finishing the experiments following the previous design, we found that the GCE n1-highcpu-2-d instance failed in running FT with workload Class B. This could be due to its relatively small size of memory. Therefore, we only show the benchmarking result with workload Class A, as shown in Fig.~\ref{fig:subfigComputation}. As claimed by Google, since individual virtual cores of different GCE VM types have the same computation power (i.e.~2.75 GCEUs) \cite{Google_instance}, the single-process performance of the evaluated four GCE instances did not reveal much difference. Nevertheless, when increasing the benchmark processes, only the high memory GCE type gave notable improvement of computation performance, which is surprising. With virtually dual-core processors, except for n1-standard-1-d, the other two GCE types should have also exhibited significant scalability from one process to two processes. 

Unlike the phenomenon on GCE, the expected rule of scalability can be clearly observed on EC2 instances. Similarly, here we only specify the experimental results of two EC2 types: m1.large and m3.xlarge. Among the evaluated EC2 instances, m1.large is the lowest-SLA EC2 type that is comparable to GCE at two-process performance (cf.~Fig.~\ref{fig:subfigComputation:e}), while m3.xlarge is the lowest-SLA EC2 type for comparison to GCE at single-process performance (cf.~Fig.~\ref{fig:subfigComputation:f}). As can be seen, multi-core EC2 instances can nearly double their computation performance when switching the benchmarks' process number from one to two. As a result, even EC2 m1.large with relatively poor single-process performance can still come up with most GCE types when running multi-process benchmarks.   

To straight portray the original scalability \cite{Li_OBrien_2012a} of different GCE and EC2 instances, we employed the typical scalability metric \textit{Speedup Over a Baseline} \cite{Li_OBrien_2012b}, as shown in Fig.~\ref{fig>PicSpeedup}. In particular, we set the baseline as the single-process performance of EC2 m1.large with the LU benchmark (recall that LU is the only pseudo application benchmark that runs on a power-of-2 number of processes \cite{NASA_2013}). 
It is then clear that multi-core EC2 instances are generally more originally-scalable than GCE instances. Moreover, although the high memory GCE type appeared to have better original scalability, its performance would become significantly variable under multi-process circumstances (cf.~Fig.~\ref{fig:subfigComputation:c}). These findings reveal that GCE may not be ready yet for single VM-based parallel computing.  

Given the current prices of the evaluated GCE and EC2 types, we further calculated their performance/price ratio, as listed in Table \ref{table_ratio}. The performance here refers to the average computation transaction speed of running LU at two processes. Interestingly, even with poor original scalability, GCE seems to have generally economic advantage over Amazon EC2. Only EC2 c1.medium beats all the evaluated GCE types in this case. This finding suggests that GCE is a relatively economic option in terms of computation, and it could be even more economic if improving GCE to be as originally scalable as EC2.

\begin{table}[!t]
%% increase table row spacing, adjust to taste
\renewcommand{\arraystretch}{1.3}
% if using array.sty, it might be a good idea to tweak the value of
% \extrarowheight as needed to properly center the text within the cells
\caption{Computation Performance/Price Ratio of Different VM Types using LU at Two Processes}
\label{table_ratio}
\centering
%% Some packages, such as MDW tools, offer better commands for making tables
%% than the plain LaTeX2e tabular which is used here.
\begin{tabular}{|l|c|c|c|}
\hline
\textbf{VM Type} & \textbf{Price} & \textbf{LU Computation} & \textbf{Performance/Price}\\
\hline
EC2 m1.medium & \$0.120/hour & 1337.34 Mop/s & 401202.71 Mop/cent \\
\hline
EC2 m1.large & \$0.240/hour & 2777.3 Mop/s & 416595 Mop/cent \\
\hline
EC2 m1.xlarge & \$0.480/hour & 2725.9 Mop/s & 204442.61 Mop/cent\\
\hline
EC2 m2.xlarge & \$0.410/hour & 3639.32 Mop/s & 319549.96 Mop/cent\\
\hline
EC2 m2.2xlarge & \$0.820/hour & 3629.46 Mop/s & 159342.33 Mop/cent\\
\hline
EC2 m3.xlarge & \$0.500/hour & 4210.94 Mop/s & 303188.03 Mop/cent\\
\hline
EC2 m3.2xlarge & \$1.000/hour & 4340.91 Mop/s & 156272.65 Mop/cent\\
\hline
EC2 c1.medium & \$0.145/hour & 2690.39 Mop/s & 667960.02 Mop/cent\\
\hline
EC2 c1.xlarge & \$0.580/hour & 2773.06 Mop/s & 172120.72 Mop/cent\\
\hline
GCE standard-1-d & \$0.132/hour & 2343.21 Mop/s & 639058.31 Mop/cent\\
\hline
GCE standard-2-d & \$0.265/hour & 2833.51 Mop/s & 384930.02 Mop/cent\\
\hline
GCE highmem-2-d & \$0.305/hour & 3787.07 Mop/s & 446997.85 Mop/cent\\
\hline
GCE highcpu-2-d & \$0.163/hour & 2836.06 Mop/s & 626368.94 Mop/cent\\
\hline
\end{tabular}
\end{table}

\section{Conclusions and Future Work}
\label{conclusion}
Cloud computing with public Cloud services has been widely regarded as a potential and encouraging paradigm for scientific computing. Since the recently available GCE is supposed to meet high-performance and computationally intensive workloads \cite{Google_instance}, we performed an early-stage evaluation of GCE to verify its fundamentally scientific capabilities. By contrasting with a set of Amazon EC2 types, the experimental results and analyses exhibit both possible advantages of, and threats to employing GCE, as listed below. 

The potential advantages of using GCE:
\begin{itemize}
\renewcommand{\labelitemi}{$\bullet$}
\itemsep 2pt
    \item	Relatively high memory data throughput. Different GCE types seem to have consistent type of virtual memory. 
    \item	Relatively fast storage transaction speed.
    \item	Relatively fast computation transaction speed with single process.
    \item	Relatively computationally economic, given generally high computation performance/price ratio.
\end{itemize}

The possible threats to using GCE:
\begin{itemize}
\renewcommand{\labelitemi}{$\bullet$}
\itemsep 2pt
    \item	Relatively low communication data throughput between US and European data centers. Distributed computing crossing both data centers may need to be well balanced.
    \item	Considerable variability of storage performance. GCE seems to have a lack of dedicated storage cache.
    \item	Considerable variability of computation transaction speed on high memory VM type.
    \item	Relatively poor (original) scalability when switching process numbers on individual VMs. GCE's hyper-thread-based virtual core seems not a suitable mechanism for parallel computing. 
\end{itemize}

Based on the outcomes of this study, our future work will be unfolded along two directions. The first direction is to supplement the fundamental evaluation of GCE. By applying more resources, we will try to evaluate the other VM types to investigate GCE's vertical scalability, and evaluate VM clusters to investigate GCE's horizontal scalability. The second direction is to deploy and evaluate scientific applications within relatively sophisticated GCE environment. The performance of the evaluated applications would then reflect the scientific capability of the entire GCE environment.


\section*{Acknowledgment}
NICTA is funded by the Australian Government as represented by the Department of Broadband, Communications and the Digital Economy and the Australian Research Council through the ICT Centre of Excellence program.

\end{document}